\documentclass[%
 aapm,
 mph,%
 amsmath,amssymb,
superscriptaddress,
prb,
preprint,%
nobibnotes,
]
{revtex4-1}
\usepackage{graphicx}% Include figure files
\usepackage{epstopdf}
\usepackage{bm}% bold math
\usepackage{caption}
\usepackage{subcaption}
\usepackage{float}
\begin{document}

%\preprint{AAPM/123-QED}

\title{Fano type transparency and other multimode interference effects in all-dielectric nanoshells}

% \author{Srishti Garg}
% \author{ Murugesan Venkatapathi*}
%  \affiliation{Computational Photonics Laboratory, SERC, Indian Institute of Science, Bangalore, 560012}
% \email{*murugesh@serc.iisc.ernet.in}

\author{Srishti Garg}
\author{Murugesan Venkatapathi}
 \email{murugesh@serc.iisc.ernet.in}
\affiliation{
Computational and Statistical Physics Laboratory, Department of Computational Science, Indian Institute of Science, Bangalore, 560012}
%  Authors' institution and/or address\\
%  This line break forced with \textbackslash\textbackslash
% }%
\begin{abstract}
Recently, the coupling of two different modes of a homogeneous plasmonic particle and their sharply varying spectra were elucidated as Fano resonances; an  'interference' of two spatially orthogonal modes driving each other.  On the other hand, the scattering (and extinction) cross-section of a non-absorbing dielectric particle is always the sum of the cross-sections of all mode numbers; and this rules out any such Fano type interference between two different mode numbers.  So delectric particles exhibit an interference structure in their extinction spectra only if it manifests in the individual modes describing the scattered field of the particle.
We show that in a \textit{all-dielectric} core-shell particle such strong interferences in multiple mode numbers can be attained, and notably even as a spectral region of transparency
and directional scattering of incident light. Here interference between the complementary normal modes of the nanoshell and core regions can be realized for each mode number, resulting in a sharp interference structure in the extinction of the particle. This manifests as spectral regions of minimal/maximal interaction with the incident electromagnetic field. Such spectral properties are significant for many applications where the non-radiative losses of plasmonic
structures are a liability. Note that this behaviour is useful for optical antennas, cloaking materials and a quantum mechanical interpretation of this
classical effect may be signicant for single-photon based applications.
\end{abstract}

%\pacs{78.20.Bh, 8.67.Bf 42.25.Hz}% PACS, the Physics and Astronomy
\maketitle

\section{Introduction}
Core-shell nanoparticles were studied extensively in recent years for high optical efficiencies, and due to sensitivity of their optical properties to geometrical parameters 
\cite{1, 2, 3}.
 In plasmonic metal nanoparticles, the electromagnetic energy can be extremely localized with apparent near and far field optical properties.  This property can be used in modification
 of local emission properties, increasing efficiency of absorption, and many biological applications \cite{4, 5, 6}.  Nevertheless the large non-radiative losses in such metal particles discourage
 their use in bulk materials for many applications.  Hence, discovering useful optical properties in non-plasmonic nanostructures is a significant direction of research in developing nanoscale
materials \cite{7}. One path in realizing new useful optical properties is the strong excitation of higher order modes of nanostructures like nanoshells 
\cite{8, 9, 10, 11}.

Studies of layered nanospheres with an absorbing layer or core were motivated by meteorological
applications many decades before the advent of plasmonics \cite{12}.  More recently, tunable properties of plasmonics
core-shell nanostructures have been studied for applications where low quality factors and dissipative losses are not an
impediment. Non-intuitive properties of magnetic spheres like a negligible back-scattering efficiency were predicted decades ago, \cite{13}
and recent experimental observations of certain cylindrical oligomers confirm such behavior \cite{14, 15, 16}.  But studies of dielectric particles 
have been mostly limited to large cylindrical and spherical microstructures for optical fiber communication and optical cavities that exploit
whispering gallery modes, respectively \cite{17, 18, 19, 20, 21}. Note that an interference structure in the extinction spectrum of homogeneous particles
on the order of wavelength has been well-known \cite{22}. This was understood as the spectral maxima and minima representing the interference of forward-scattered light
and incident light, and this effect has been used in micro-wave signal processing before\cite{23}. Also, an asymmetric optical resonance for
particles was predicted by Van de Hulst earlier than its quantum-mechanical counterpart\cite{24} and such a resonance was experimentally observed using nanospheres with
a sharp absorption spectrum \cite{25} before Fano resonances became well-known in plasmonics.  Plasmonic particles offer the possibility of sharp changes in the optical spectra when two different mode numbers are coupled strongly by the dissipative currents.

However, we highlight that strong optical effects can be derived even in dielectric nanoshells with dimensions on the order of, but less than,
wavelength of incident light \textit{i.e} 2$\pi$ $>$ \textit{kd}\ $>$ 1, where \textit{k} is the magnitude of incident wave-vector and \textit{d} is the dimension of layers.
Spherical nanoshells in particular allow us to derive polarization independent resonances.
Transparencies where the optical efficiency of a large nanoparticle reduces to unity due to a minimal interaction with the incident field are equally significant spectral regions \cite{26, 27}.
We show that interference occur between the complementary normal modes of the nanoshell and core regions, due to large phase shifts relative to each other, in many mode numbers. 
In other spectral regions, the optical properties are not remarkably different from that of dielectric homogeneous spheres \cite{21, 29}.
Note that this is accompanied by directional scattering typically in forward or backward directions and hence a useful property for optical antennas\cite{28}.
This classical behavior has interesting quantum-mechanical interpretations as discussed later, and is potentially conducive to single-photon communication applications.
We describe the optical properties of nanoshell particles using the coefficients of the spherical modes in the well-known Lorenz-Mie theory \cite{30}; the magnitude and phase of
mode coefficients $a_{n}$, $b_{n}$ for the external scattered field; $c_{n}$, $d_{n}$, for the normal modes of the core region;  $f_{n}$, $g_{n}$, $v_{n}$, $w_{n}$
for the normal modes of shell region. The properties of scattered field and resulting absorption/scattering are typically studied using mode coefficients $a_{n}$, $b_{n}$. We show that
the origin of such effects come from the interference of the complementary electric and magnetic normal modes of the core-shell region. This requires a study of the higher order
internal modes of the core-shell particle using all the above mentioned coefficients.  We studied nanospheres with silica as dielectric core which
is very stable in nature \cite{31, 32}, and a shell of titania which is an unremarkable non-absorbing material in its VIS properties, though strongly absorbing in the UV 
\cite{33, 34}.
We show that hollow titania nanoshells also exhibit similar behavior in the absence of the silica core.
Numerical results also suggest that these interference effects can survive size-dispersion in actual realization of materials while allowing
spectral tunability based on the nominal geometrical parameters. In case of larger particles both resonances and transparencies manifest at different spectral regions of the
particle and these are broad.

\section{ANALYTICAL MODEL}

Consider a linearly polarized electromagnetic wave  incident on a coreshell nanosphere with inner radius `a' and outer radius is `b' (Figure 1). The incident
electric vector is polarized in the direction of the x axis and the direction of propagation of the incident wave is along the positive z axis. In spherical coordinates,
if the amplitude of the incident wave at the origin is $E_{0}$, then the incident field $E_{i}$, $H_{i}$, electromagnetic field in the core region $E_{1}$, $H_{1}$, scattered field
 $E_{3}$, $H_{3}$ and the electromagnetic field in the shell region $E_{2}$, $H_{2}$ can be expressed in spherical harmonics \cite{22, 29}(see appendix).
Where $a_{n}$ and $b_{n}$ denote the coefficients of  scattered field, $f_{n}$, $g_{n}$, $v_{n}$, $w_{n}$ represents the coefficients of 
normal mode in shell region and $c_{n}$, $d_{n}$ the coefficients of normal mode core region respectively (refer Eq(5-12) in appendix).

\begin{figure}[H]
 \centering
\includegraphics[width=0.5\linewidth,height=0.32\textheight]{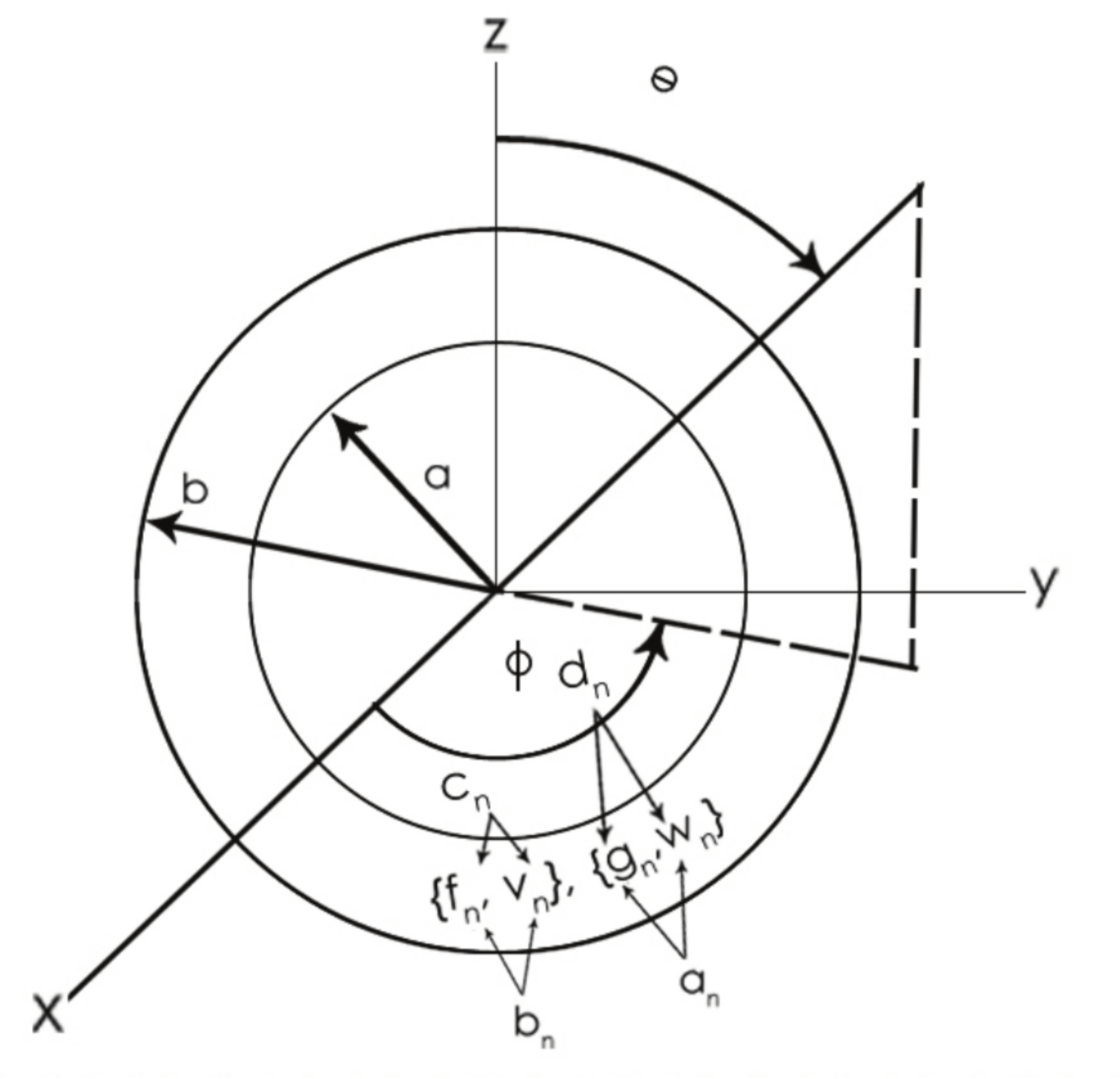}
\caption{Schematic of the coordinate system used}
\end{figure}

The extinction and scattering cross-sections are given by
 \begin{equation}
 \ C_{ext}= \frac{2\pi}{k^2}\sum_{n=1}^{\infty}(2n+1) \mathbb{R}(a_{n}+b_{n})
 \end{equation}

 \begin{equation}
\ C_{sca}= \frac{2\pi}{k^2}\sum_{n=1}^{\infty}(2n+1)(|a_{n}|^2+|b_{n}|^2)
 \end{equation}

 \begin{equation}
\ C_{abs}=  C_{ext} - C_{scat}
 \end{equation}
 Also efficiency $Q=\frac{C}{\pi r^2}$\\

The efficiency of the backscattering $Q_{b}$ is
 \begin{equation}
 \ Q_{b}= \frac{1}{4\pi}\bigg[\frac{1}{k^2 R^2} \bigg|\sum_{n}^{}(2n+1)(-1)^n(a_{n}-b_{n})\bigg|^2\bigg]
 \end{equation}
 where `R' is the radius of the sphere; note the factor `4 $\pi$' include here which means that the differential scattering cross-section at $\theta =180^{\circ}$ is used to calculate $Q_{b}$.
 This additional factor is sometimes ignored to express relative backscattering cross-section scaled to represent the total surface area of the particle \cite{22}.

The final solutions for mode coefficients $a_{n}$, $b_{n}$, $c_{n}$, $d_{n}$, $f_{n}$, $g_{n}$, $v_{n}$, $w_{n}$ have been derived
as in Equations (19-26) (see appendix).
The numerical evaluations of cross-sections using the above Lorenz-Mie formalism \cite{22, 35, 36} were also verified and complemented by numerical volume integral 
methods such as discrete dipole approximation. \cite{37} All the simulations were made with water, as an ambient medium of permittivity 1.77. These results of internal polarization are
also used in elucidating the physics in the next section.  As given by Eq (5-6), we studied nanoshells illuminated by plane electromagnetic waves.
We assume the titania is of anatase phase and the dispersive permittivity of anatase is different from its rutile phase \cite{33, 34, 38, 39}
(see supplementary material for the refractive indices of the materials used).

 \section{RESUTS AND DISCUSSION}
%\floatstyle{capposition=top}
% \floatsetup[figure]{style=plain }
%\captionsetup[subfigure]{labelformat=parens, labelfont=small}
%\captionsetup{position=top}

Figure 2 shows the extinction spectra of silica-titania spheres of three increasing dimensions with varying ratios of thickness of shell and radius of the particle.
As mentioned before, the larger shells exhibit non-intuitive spectra which we show later to be due to the interference between higher order modes.
The 180 nm dia particles (represented in Figure 2a)  are unremarkable as they show a Rayleigh scattering type $\frac{1}{\lambda^{4}}$ behavior for $\lambda>$ 450 nm as expected,
followed by an indication of interference structure around the optical efficiency of 2 for 400 nm $<\lambda<$ 450 nm.  The higher energies ($\lambda<$ 350 nm) are strongly absorbed by
titania and the extinction spectra show a saturation behavior based on thickness of the shells \cite{40}.
The 360 nm dia particles (represented  in Figure 2b) exhibits a similar unremarkable Mie-Rayleigh behavior for wave-lengths larger than 500 nm based on its larger size.

\begin{figure}[H]
    \centering
    \begin{subfigure}[b]{0.35\textwidth}
 %       \centering
    \includegraphics[width=1\linewidth,height=0.21\textheight]{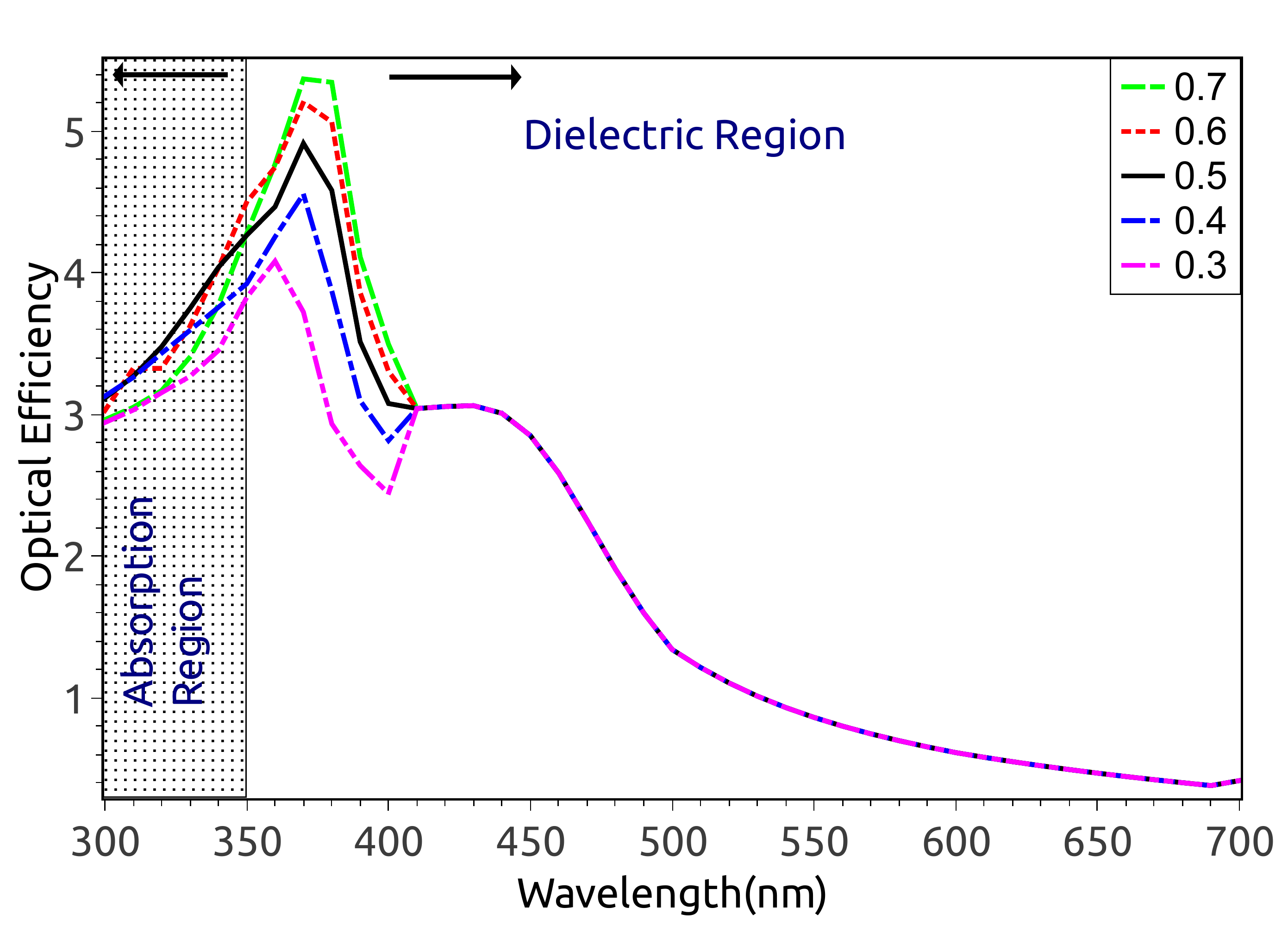}
        \caption{}
    \end{subfigure}%
 ~   
    \begin{subfigure}[b]{0.35\textwidth}
 %       \centering
      \includegraphics[width=1\linewidth,height=0.21\textheight]{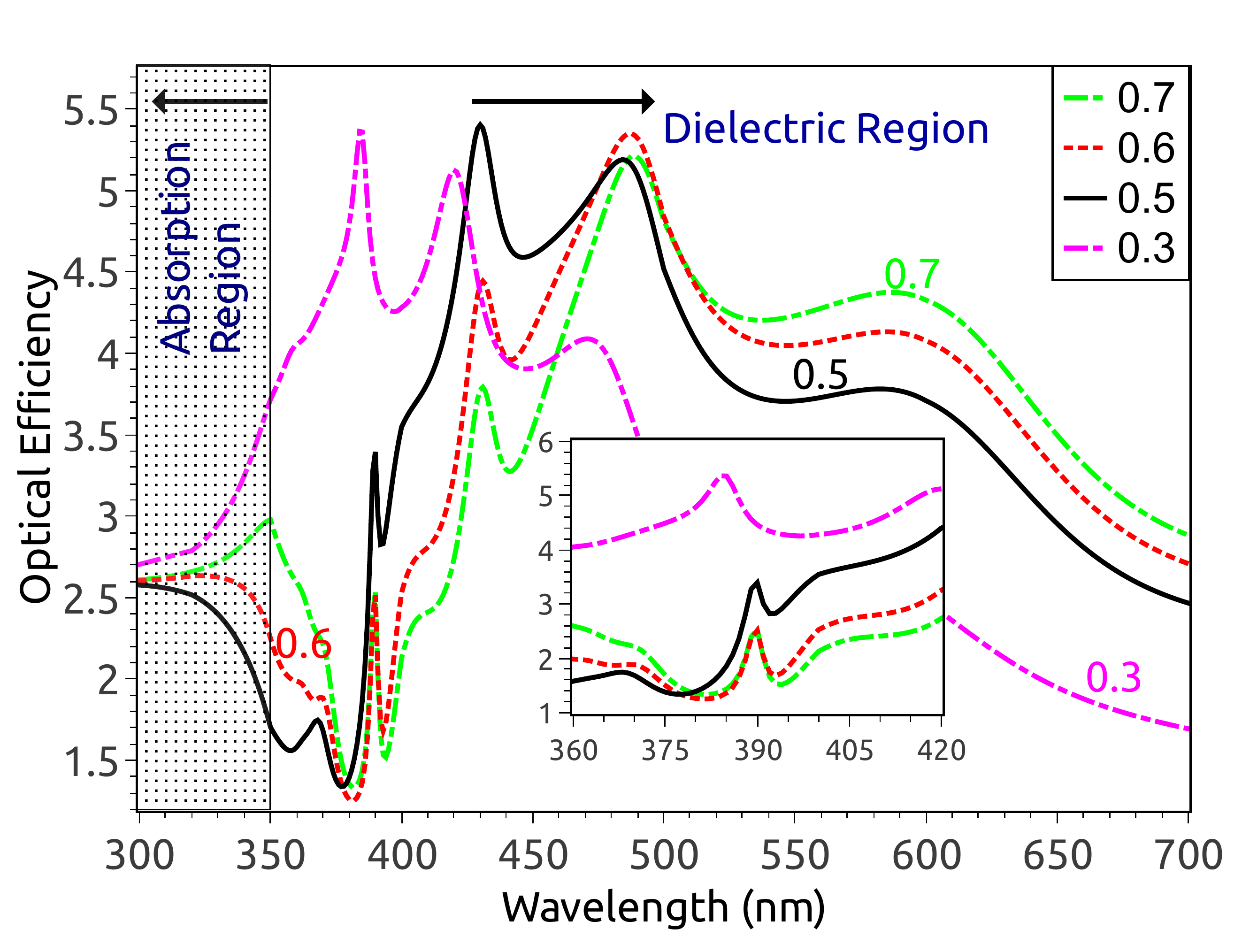}
        \caption{}
    \end{subfigure}%
~   
     \begin{subfigure}[b]{0.35\textwidth}
 %       \centering
     \includegraphics[width=1\linewidth,height=0.21\textheight]{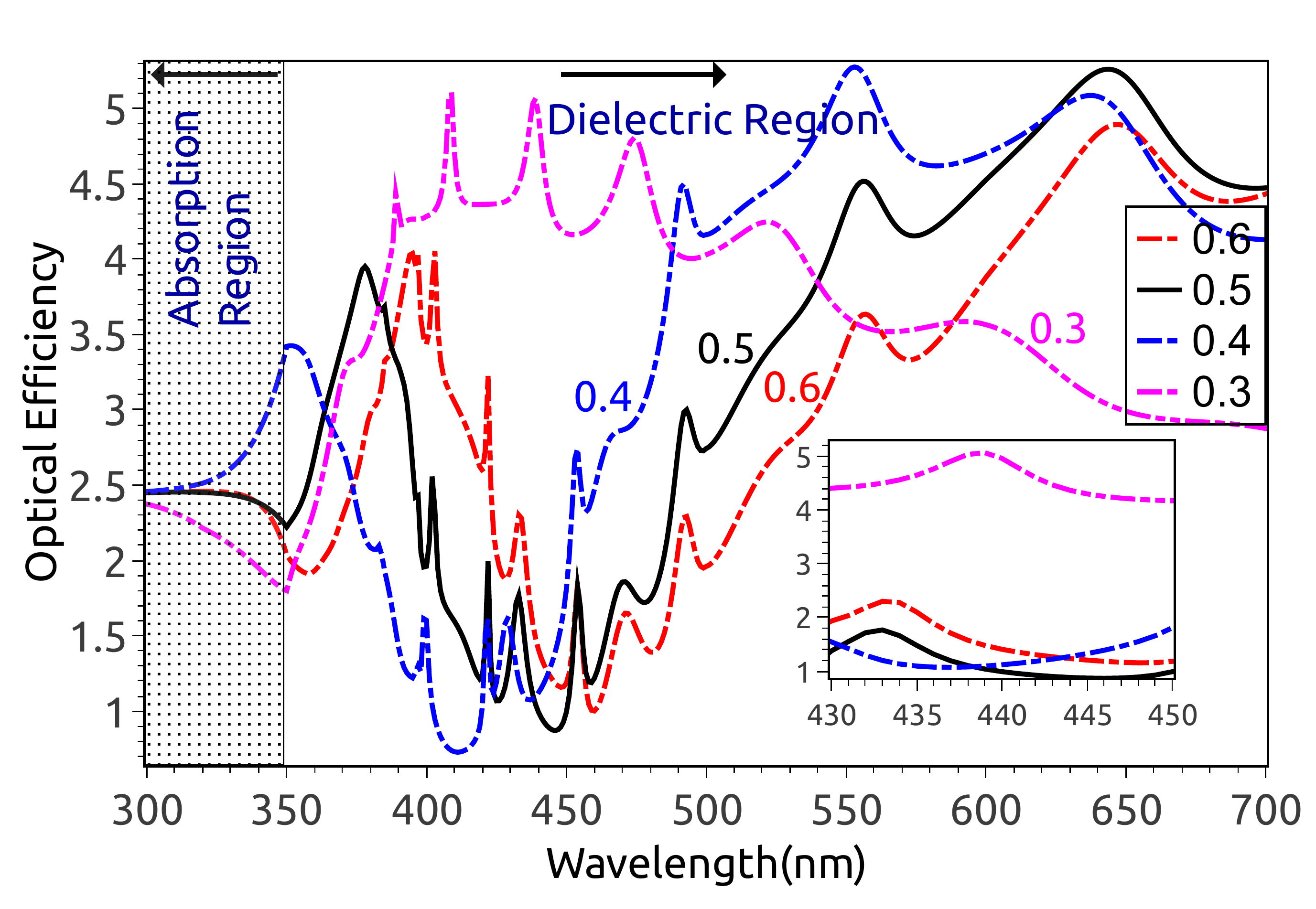}
        \caption{}
    \end{subfigure}
    
   \caption{Extinction spectra as a function of freespace wavelength for silica-titania spheres at different shell ratios 0.3 to 0.7
            (given by the ratio of the thickness of shell and outer radius of particle);
             (a) 180 nm diameter.    (b) 360 nm diameter.   (c) 600 nm diameter.
             Note that the similar extinction spectra for hollow shell particles are presented in supplementary material.}
  \end{figure}
  \vspace{-20em} 
On the other hand, an abrupt reduction in its optical efficiency to $\sim$ 1 for the spectral range 350-430 nm range is observed (see appendix for definition of efficiencies).
This efficiency is 5 times smaller than that of the smaller 180 nm sphere of similar constitution, and note that while both titania and silica are in fact mostly non-absorbing in this
energy spectrum, the lack of stronger scattering for the larger particles seems anomalous. Weak absorption close to the absorption edge at 350 nm significantly enhances the typical
interference structure in the extinction spectrum of a dielectric particle, resulting in this sharp asymmetric resonance around 370 nm. A similar more broadband effect for the larger
sphere of 600 nm dia (represented in Figure 2c) is observed, which exhibits both a broader transparency (at $\lambda$ $\sim$ 450 nm) and another resonance (at $\lambda$ $\sim$370 nm). The tunability of this transparency on the 
size and the core-shell ratio of the particle are highlighted in these results.

\begin{figure}[H]
    \centering
    \begin{subfigure}[b]{0.35\textwidth}
        \centering
    \includegraphics[width=1\linewidth,height=0.15\textheight]{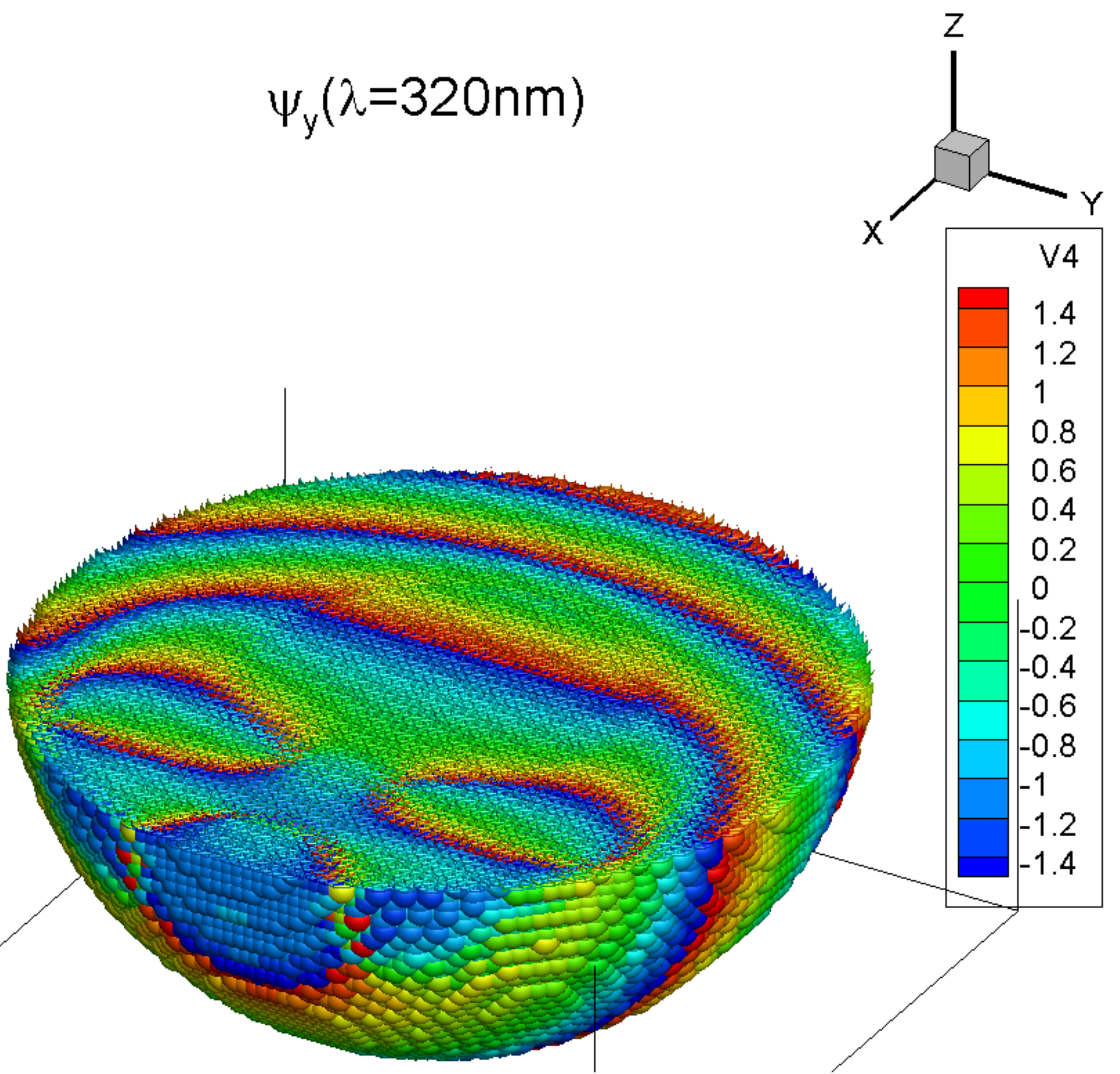}
        \caption{}
    \end{subfigure}%
    ~ 
    \begin{subfigure}[b]{0.3\textwidth}
        \centering
     \includegraphics[width=1\linewidth,height=0.12\textheight]{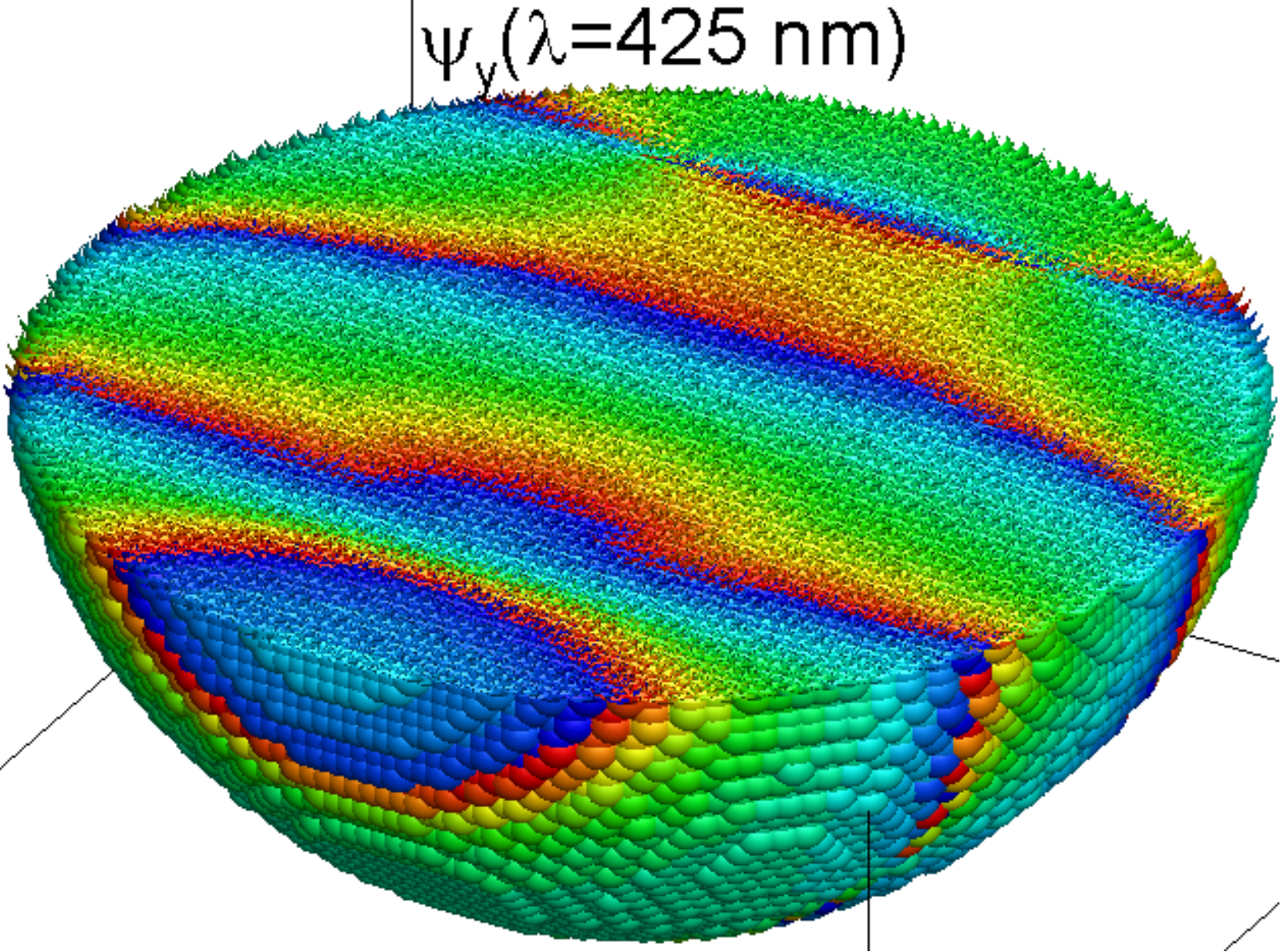}
        \caption{}
    \end{subfigure}%
    ~
     \begin{subfigure}[b]{0.3\textwidth}
        \centering
    \includegraphics[width=1\linewidth,height=0.12\textheight]{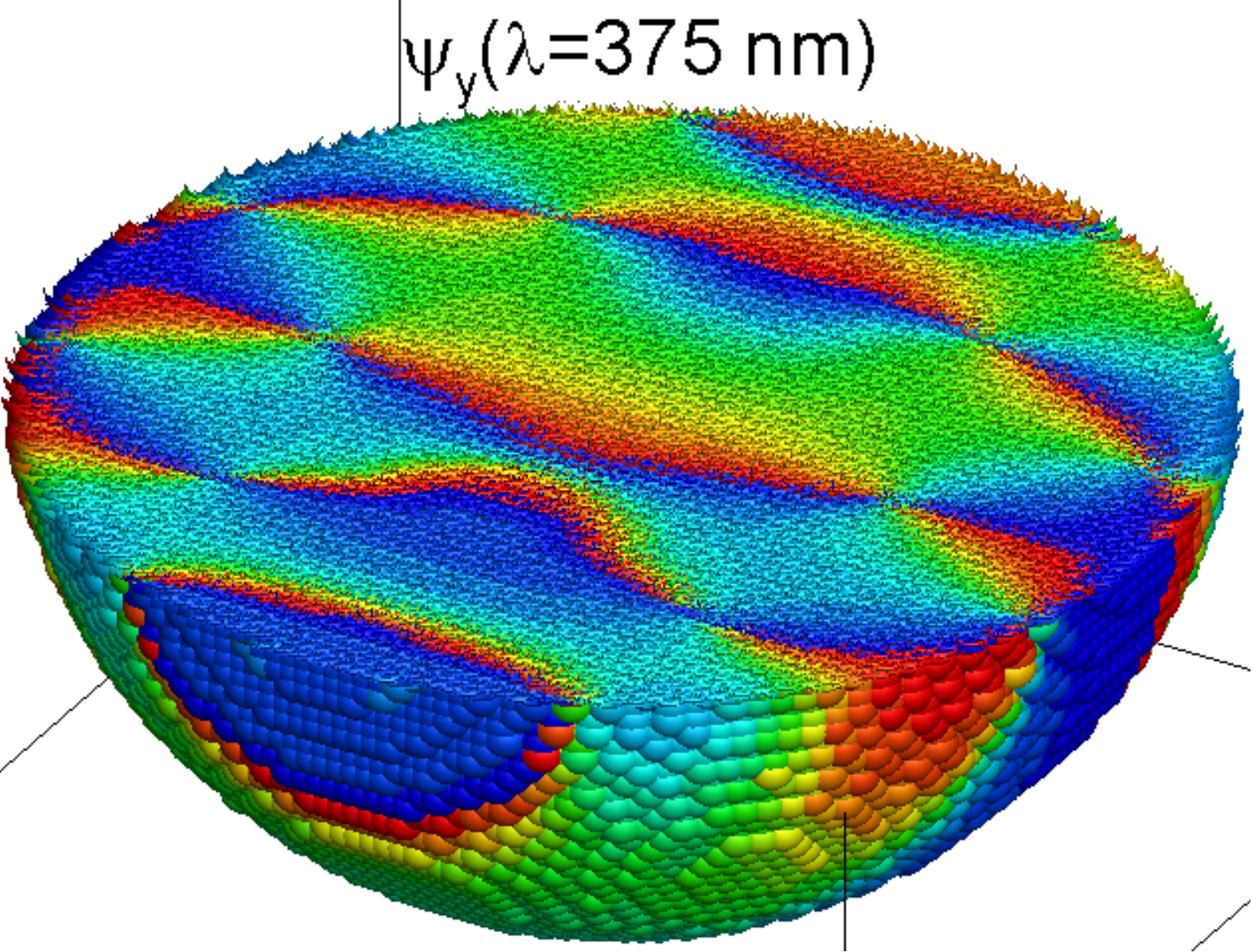}
        \caption{}
    \end{subfigure}
    
     \begin{subfigure}[b]{0.35\textwidth}
        \centering
      \includegraphics[width=1\linewidth,height=0.12\textheight]{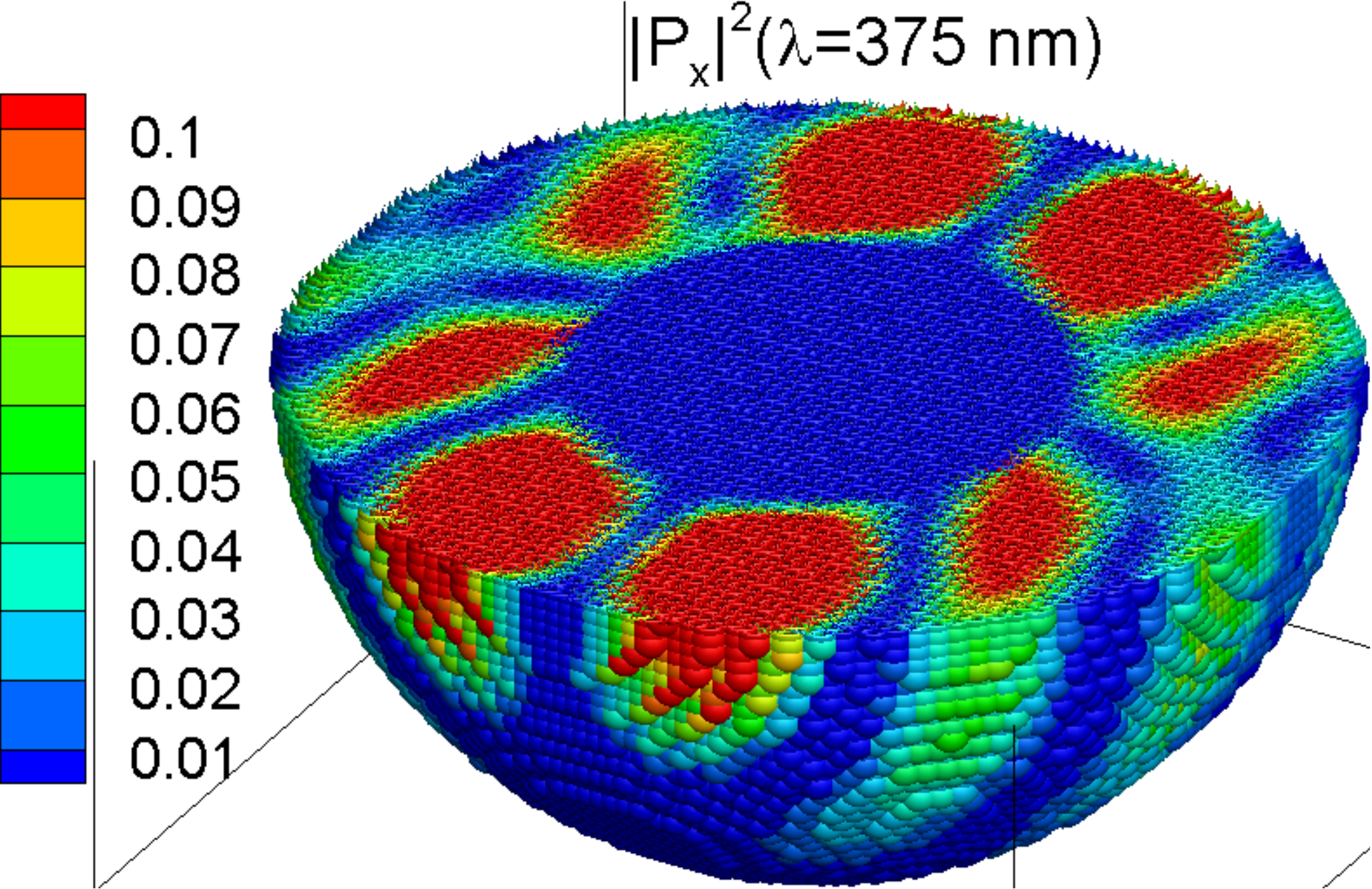}
        \caption{}
    \end{subfigure}%
    ~
     \begin{subfigure}[b]{0.3\textwidth}
        \centering
      \includegraphics[width=1\linewidth,height=0.12\textheight]{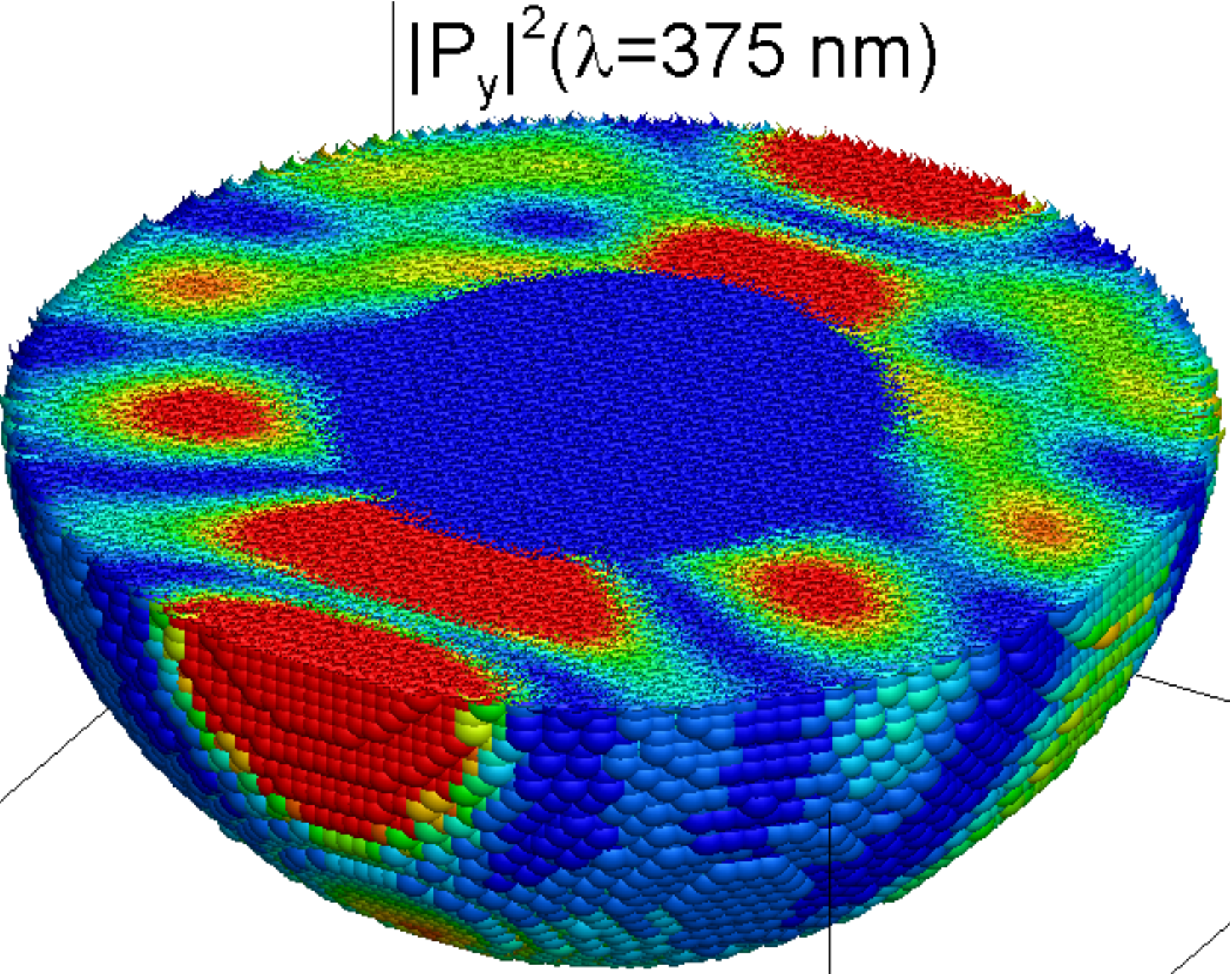}
        \caption{}
    \end{subfigure}%
    ~
     \begin{subfigure}[b]{0.3\textwidth}
        \centering
      \includegraphics[width=1\linewidth,height=0.12\textheight]{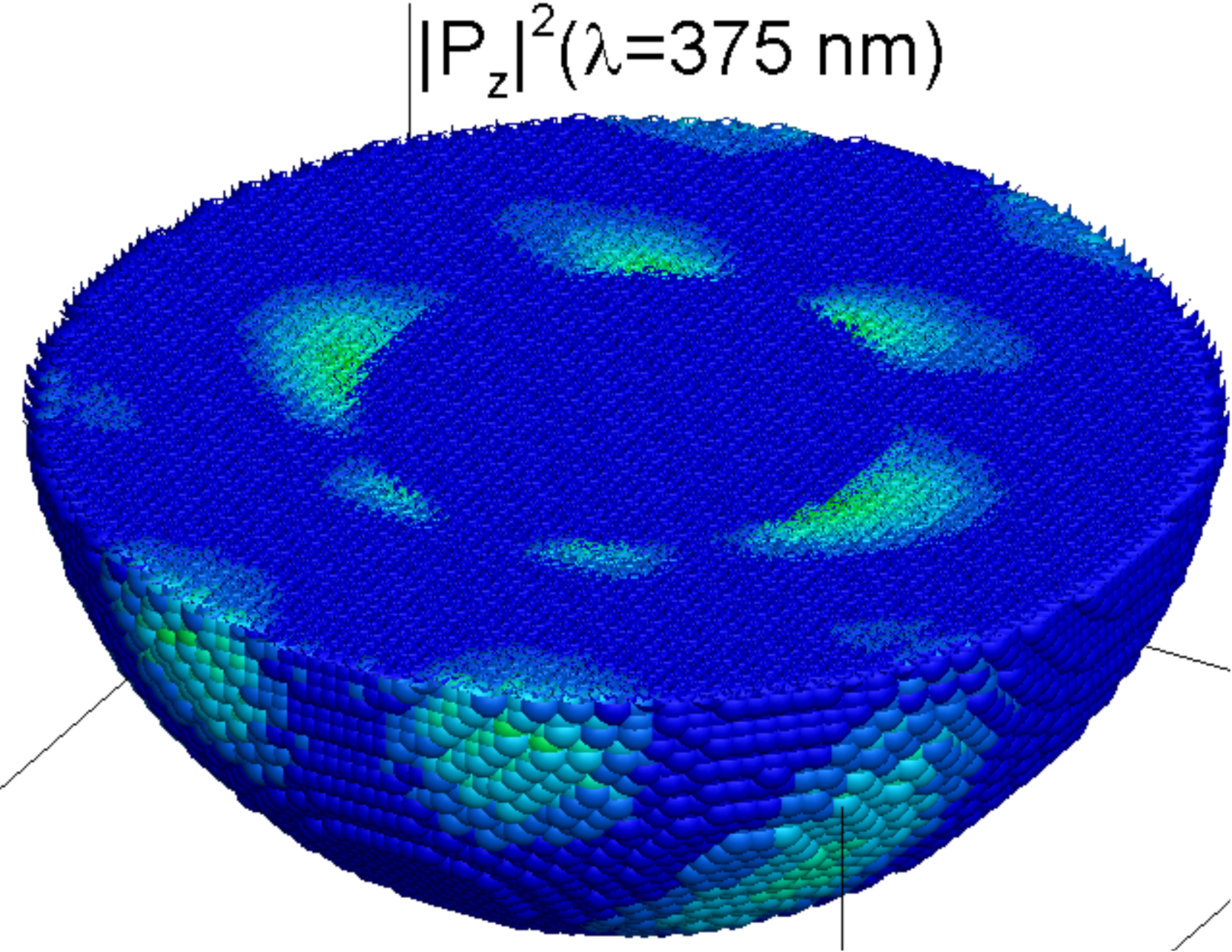}
        \caption{}
    \end{subfigure}
    \caption{Phase of polarization $\psi$ (in radians) and relative energy distribution in a 360 nm silica-titania coreshell particle of 0.5 shell ratio.
    The nanoshell particle is sectioned at Y= 0 plane to display a hemisphere and its internal distributions of polarizations.
          (a) Phase distribution of $\hat{P_{y}}$ at $\lambda$ = 320 nm.  (b) Phase distribution of $\hat{P_{y}}$ at $\lambda$ = 425 nm.
          (c) Phase distribution of $\hat{P_{y}}$ at critical $\lambda$ = 375 nm. 
          At $\lambda$ = 375 nm: (d) $|P_{x}|^2$. (e) $|P_{y}|^2$. (f) $|P_{z}|^2$.
          Note that the similar phase of polarization $\psi$ (in radians) and relative energy distribution for 600 nm nanosphere are presented in supplementary material.}    
\end{figure}
 
For a more detailed investigation, we present further numerical results of the nanoshells where thickness of the shell is half of the radius of the
particle. In fact, the phase of the internal polarizations of the core-shell structure shown in Figure 3 indicates a non-trivial effect. At higher($\lambda$= 320 nm) and lower
($\lambda$= 425 nm) energies the wave nature of the internal polarization is apparent Fig (3a and 3b); but at the critical wavelength of $\sim$ 370 nm an 
interference structure can be inferred by large differences in phase between the core and shell regions (Figure 3c).  Similar effects in plasmonic metamaterial particles have been observed,\cite{27} and the corresponding energy distributions in Figure (3d-3f) also elucidates this effect.

\begin{figure}[H]
    \centering
    \begin{subfigure}[b]{0.5\textwidth}
        \centering
    \includegraphics[width=1\linewidth,height=0.25\textheight]{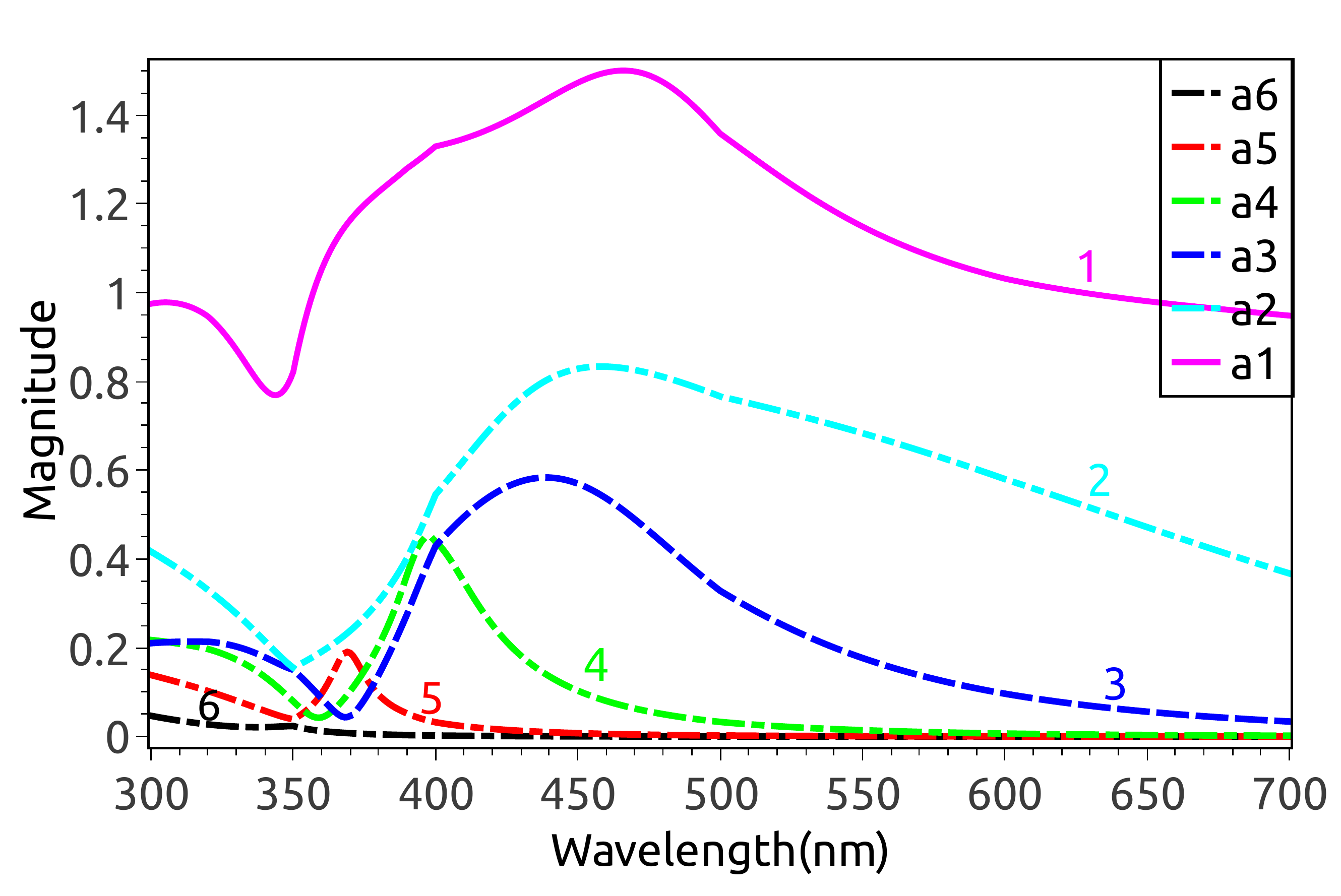}
        \caption{}
    \end{subfigure}%
    ~ 
    \begin{subfigure}[b]{0.5\textwidth}
        \centering
      \includegraphics[width=1\linewidth,height=0.25\textheight]{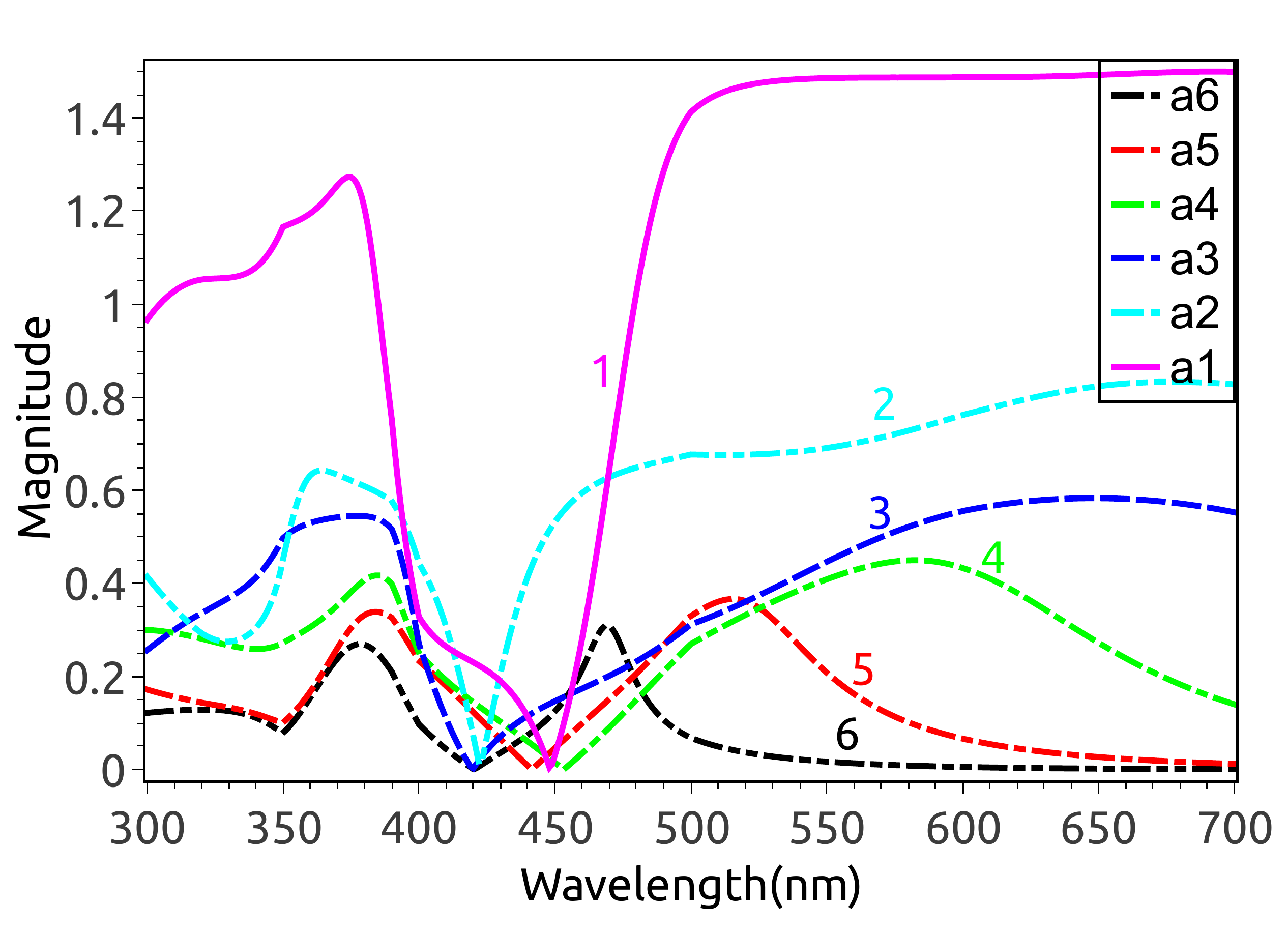}
        \caption{}
    \end{subfigure}
    ~
     \begin{subfigure}[b]{0.5\textwidth}
        \centering
     \includegraphics[width=1\linewidth,height=0.25\textheight]{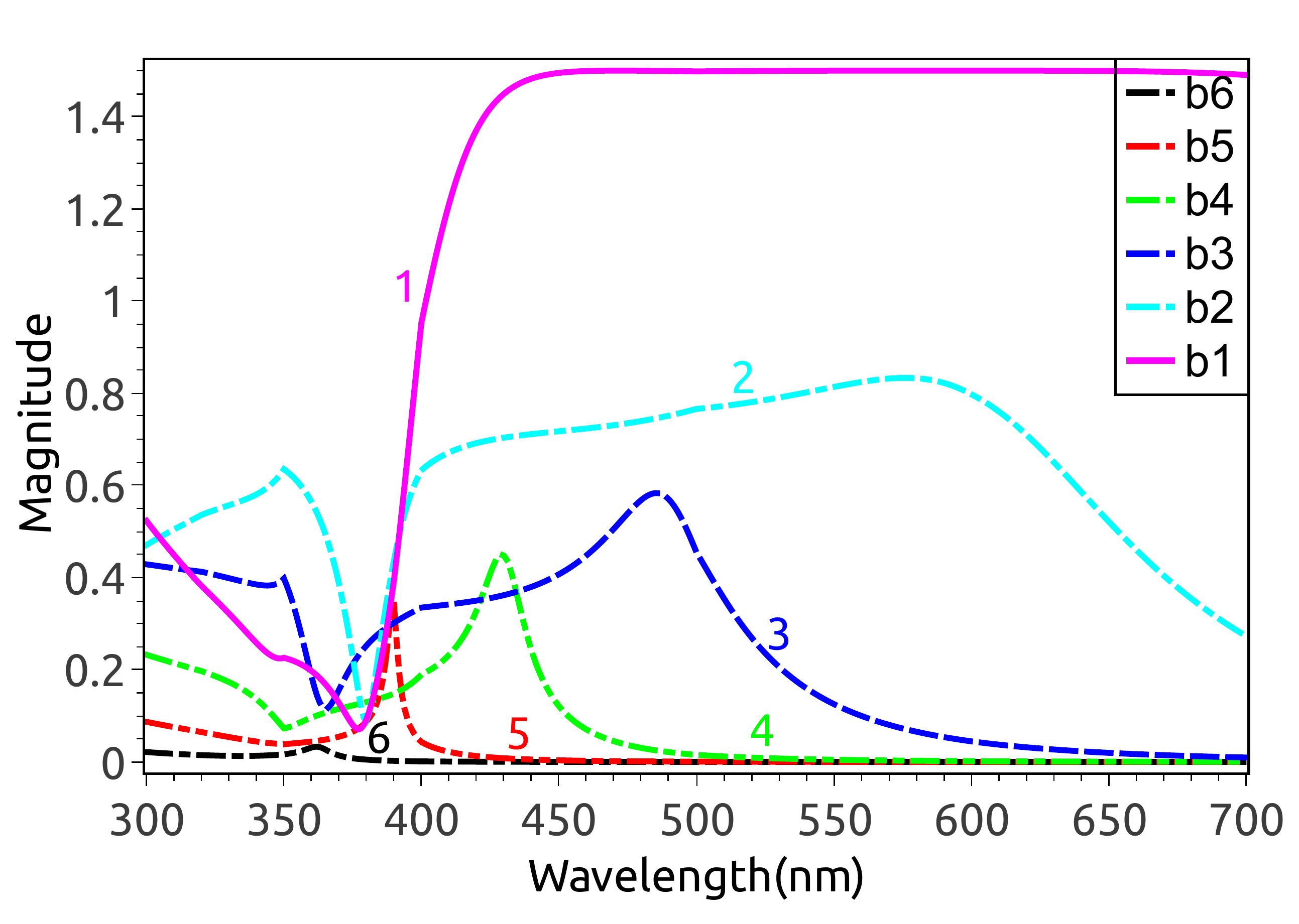}
        \caption{}
    \end{subfigure}%
    ~
     \begin{subfigure}[b]{0.5\textwidth}
        \centering
      \includegraphics[width=1\linewidth,height=0.25\textheight]{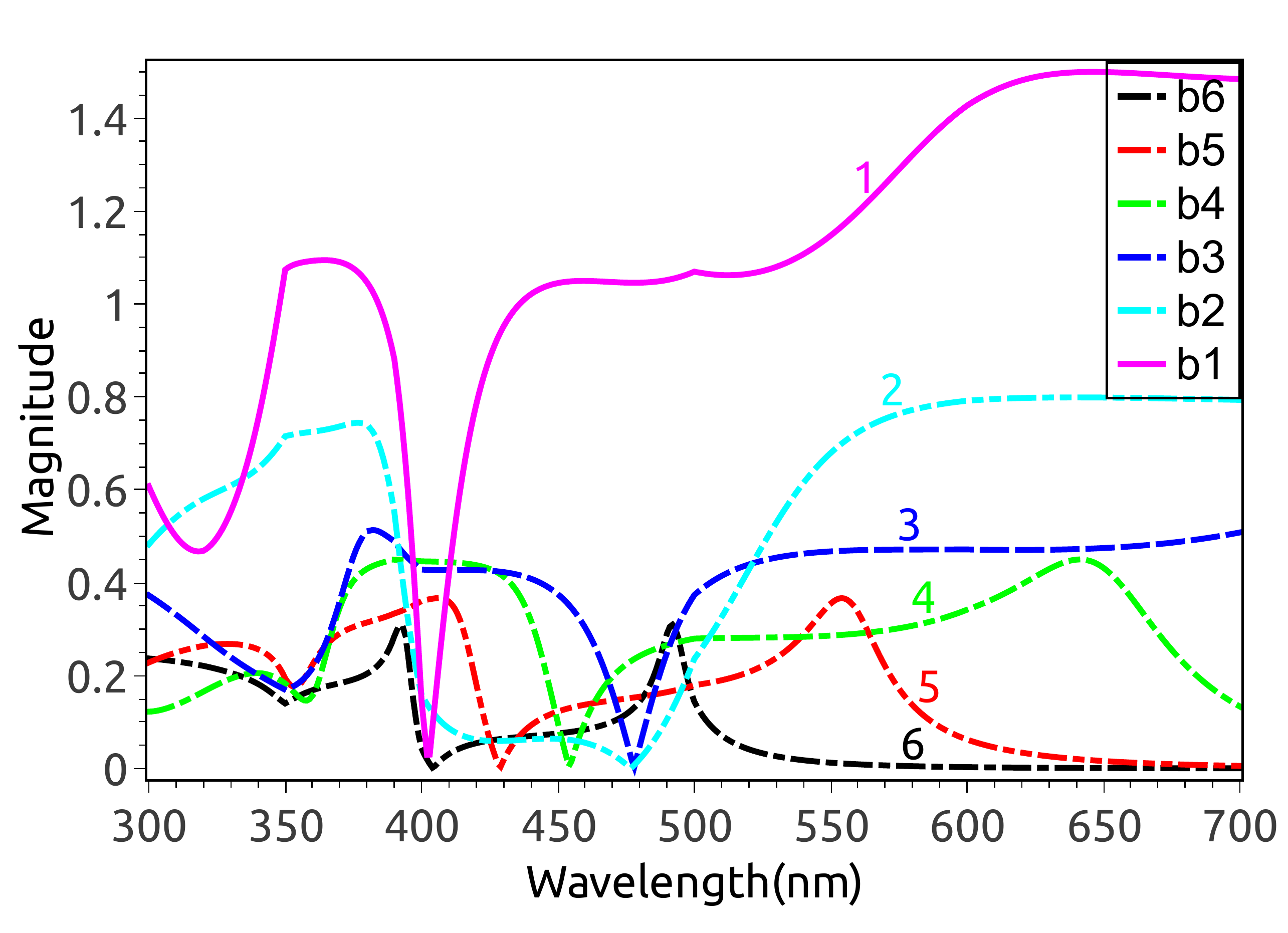}
        \caption{}
    \end{subfigure}
    
    \caption{(a-d) Normalized distribution of magnitude of spherical mode coefficients ($E_{n}*a_{n}$, $E_{n}*b_{n}$) for 360 nm (to your left side) and 600 nm (to your right side) diameter 
             particle of 0.5 shell ratio, as a function of freespace wavelength,
             where $E_{n}=\frac{2n+1}{n(n+1)}$.
             Magnitude of other mode coefficients  $b_{n}, d_{n}, g_{n}, w_{n}$ and higher modes of $a_{n}$ and $f_{n}$ (for \textit{n}= 7 to 12) 
             are plotted separately in supplementary material for clarity.}       
 \end{figure}
 
The magnitude of the spherical mode coefficients $a_{n}$, $b_{n}$ plotted in Figure 4 emphasizes the significance on the higher order non-dipolar modes, especially for larger nanoshells of 600 nm dia.
This is evident from the magnitude of the normal mode coefficients numbered 2-12; the dipolar modes are represented by coefficients numbered 1.
 An analysis of mode constants $a_{n}$ and $b_{n}$ shows that large changes in the magnitude occurs in many modes in the same regions of the spectrum.
A cumulative effect of these large changes are indeed reflected in the extinction/scaterring spectra in Figure 2.
The origin of this interference structure in a nanoshell particle requires further analysis of the internal TE and TM modes of the shell and core region. 
While $a_{n}$ and $b_{n}$ denote the normal modes for the scattered field, $f_{n}$, $g_{n}$, $v_{n}$, $w_{n}$ represents the transverse electric (TE) and magnetic (TM) modes of the shell
region and $c_{n}$, $d_{n}$ the core region respectively (refer Eq(5-12)).
The distribution of all the mode coefficients in terms of their magntiude and phase are provided in the supplemntary information.

\begin{figure}[H]
    \centering
    \begin{subfigure}[b]{0.42\textwidth}
        \centering
    \includegraphics[width=1\linewidth,height=0.2\textheight]{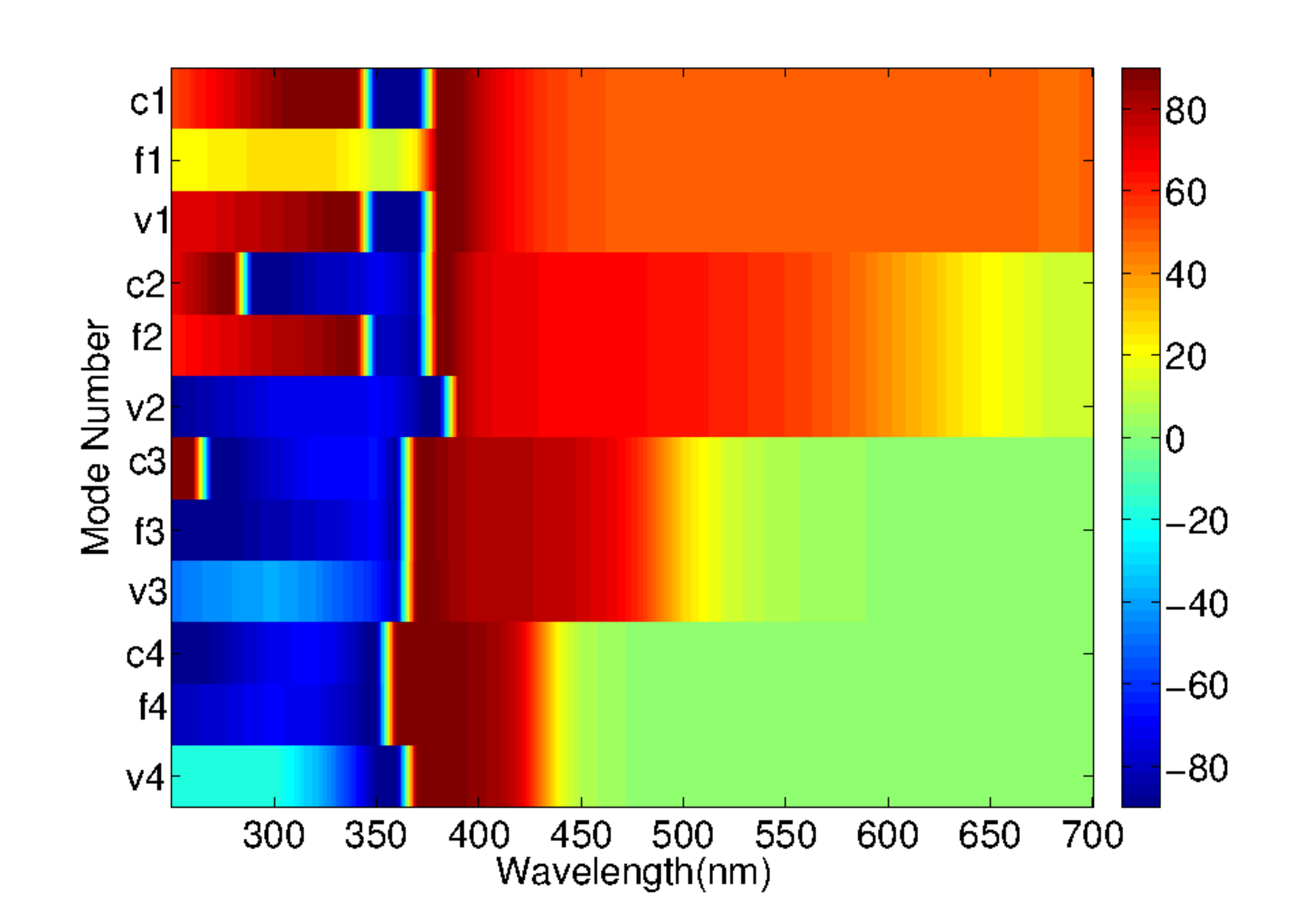}
        \caption{}
    \end{subfigure}%
~
    \begin{subfigure}[b]{0.35\textwidth}
        \centering
      \includegraphics[width=1\linewidth,height=0.2\textheight]{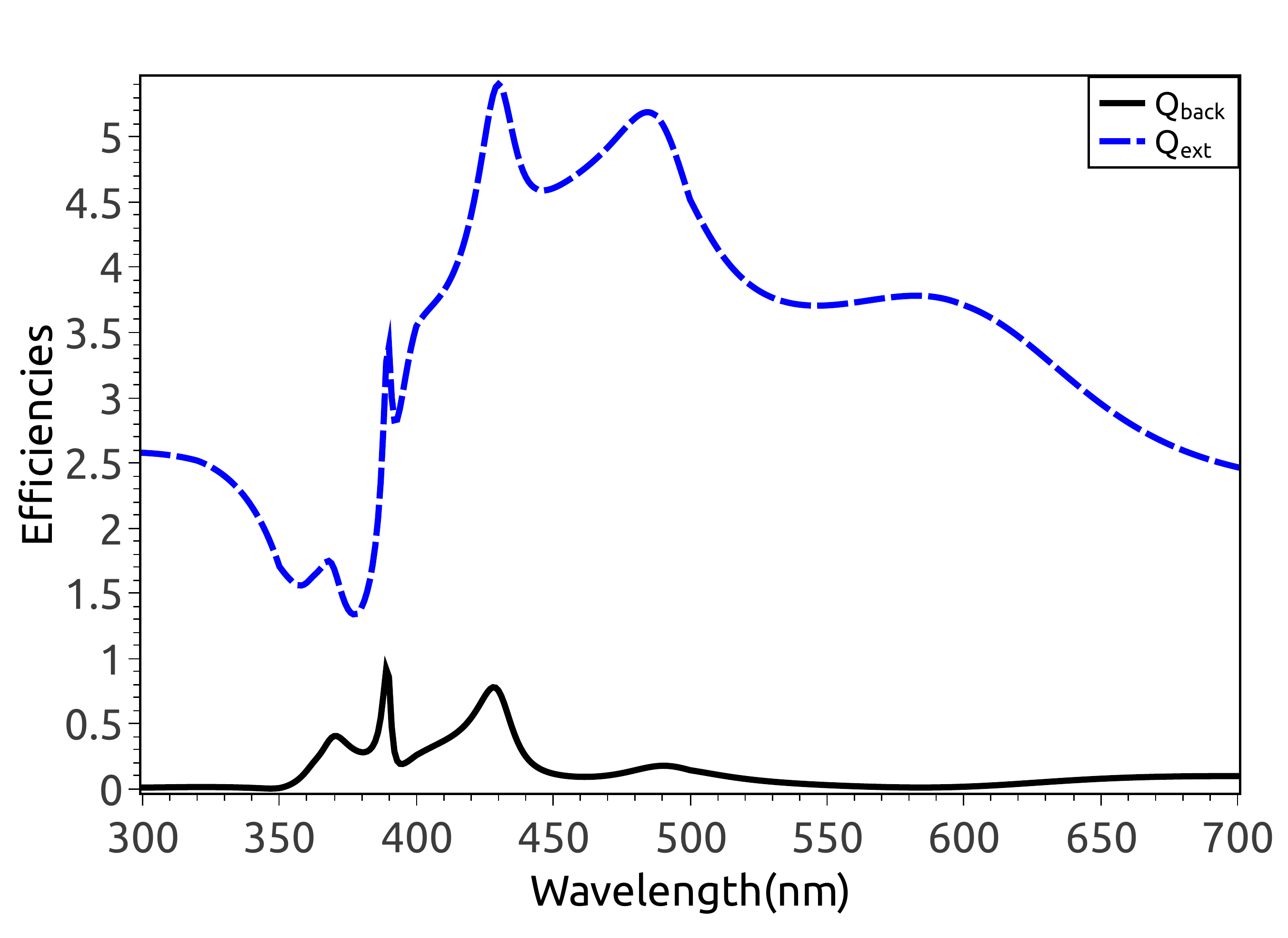}
        \caption{}
    \end{subfigure}%
~
     \begin{subfigure}[b]{0.35\textwidth}
        \centering
     \includegraphics[width=1\linewidth,height=0.2\textheight]{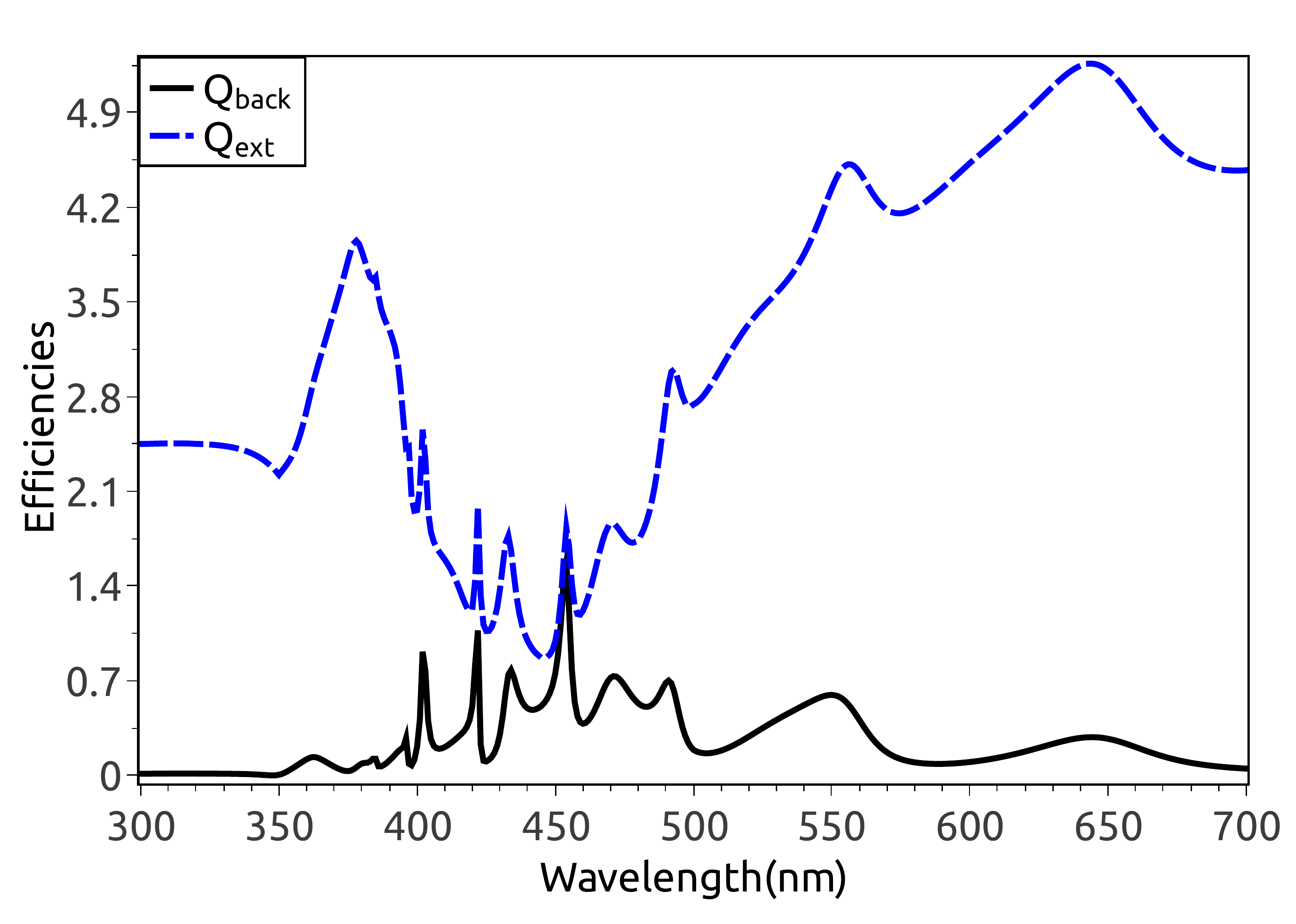}
        \caption{}
    \end{subfigure}
    
   \caption{(a) shows phase distribution (in degrees) of the spherical mode coefficients $c_{n}$, $f_{n}$ and $v_{n}$, for \textit{n}= 1 to 3 for the 360 nm diameter particle of 0.5 shell ratio,
           as a function of freespace wavelength. The 12 discrete colour strips given in figure represents 12 different mode coefficients.
           Phase distribution of other mode coefficients are given in supplementary material. (b) Exticntion and Back-scattering efficiency spectra of the the 360 nm diameter particle of 0.5 shell ratio.
(c) Exticntion and Back-scattering efficiency spectra of the the 600 nm diameter particle of 0.5 shell ratio.}
%\vspace{-1.5em} 
 \end{figure}
 
The relative phase of the spherical mode coefficients plotted in Figure 5a is illuminating.  Note that $c_{n}$, $f_{n}$ and $v_{n}$ are the complementary modes of the shell and the core that are related to the spectra of $a_{n}$.
$c_{n}$ represents the incoming and outgoing TM waves in the core, $f_{n}$ represents the incoming waves while $v_{n}$ represents outgoing electromagnetic waves of the TM modes.  The other set $d_{n}$, $g_{n}$ and $w_{n}$ are similarly related to the spectra of $b_{n}$ and TE modes. Note that unlike in the whispering gallery modes of much larger spheres, there are large phase shifts in these coefficients relative to each other in a small region of the energy spectrum.  The resulting interference between these modes of shell and core regions manifest as large changes in the magnitudes of all scattering coefficients $a_{n}$ and $b_{n}$ in a small region of the energy spectrum.  This phase behavior of complementary normal modes around the critical point of transparency, in multiple mode numbers, would allow a photon to have non-negligible probabilities to exist in multiple angular momentum states of core and shell region simultaneously.  This can result in a marked change
 of a classical system to allow coherence properties of quantum systems when the relevant dimensions are approximately $\lesssim$ $\lambda$.
 Specifically, the maximization of back-scattering component in the extinction at  $\lambda$ $\sim$ 450 nm and the forward scattering component at $\lambda$ $\sim$ 370 nm are noteworthy
 for the 600 nm particle (refer Eq. 4 in the analytical model for explanation of backscattering efficiency).  These efficiencies are  plotted in Figures 5b and 5c for the particles of the two different sizes and shell ratio of 0.5.  In the dipolar limit, the back-scattering in such an interference equals the total scattering cross-section\cite{26}.

% \begin{figure}[H]
%     \centering
%     \begin{subfigure}[b]{0.4\textwidth}
%         \centering
%     \includegraphics[width=1\linewidth,height=0.21\textheight]{6a}
%         \caption{}
%     \end{subfigure}%
% ~
%     \begin{subfigure}[b]{0.35\textwidth}
%         \centering
%       \includegraphics[width=1\linewidth,height=0.2\textheight]{6b}
%         \caption{}
%     \end{subfigure}%
%  ~   
%      \begin{subfigure}[b]{0.35\textwidth}
%         \centering
%      \includegraphics[width=1\linewidth,height=0.2\textheight]{6c}
%         \caption{}
%     \end{subfigure}
%     
%     \caption{(a) Scattering efficiencies and magnitude of Poynting vectors as a function of freespace wavelength for the 600 nm diameter particle of 0.5 shell ratio.
%           Phasor Diagrams:
%          (b) Interference structure of normal modes in the core, ($c_{n}$-i$d_{n}$) for $\lambda$ = 370 and 450 nm and
%          (c) Interference structure of normal modes in the shell, ($f_{n}$-i$g_{n}$+$v_{n}$-i$w_{n}$) for $\lambda$ = 370 and 450 nm for \textit{n}=1, 2, 3....12.}
%  % \vspace{-1.5em}    
%  \end{figure}
%  
 To summarize, the interference resonances of homogeneous dielectric spheres simply understood as interference of forward/backward scattered and the incident light becomes a richer
 phenomenon in nanoshells. The strong interference of the complementary TE and TM modes of shell and core regions of the all-dielectric particles (for every mode number) provides an alternate avenue for achieving strong optical
 properties in nanoscale materials without the absorption accompanying plasmonic nanostructures.
 
\section*{APPENDIX}

Continuing from section II, the vector spherical harmonics are given by

\begin{equation}
E_{i}=E_{0}\sum_{n=1}^{\infty} i^{n}\frac{2n+1}{n(n+1)}(M^{(1)}_{o1n}-iN^{(1)}_{e1n})
\end{equation}
\begin{equation}
H_{i}=\frac{-k}{w\mu}E_{0}\sum_{n=1}^{\infty} i^{n}\frac{2n+1}{n(n+1)}(M^{(1)}_{e1n}+iN^{(1)}_{o1n})
\end{equation}
\begin{equation}
E_{1}=\sum_{n=1}^{\infty}E_{n}(c_{n}M^{(1)}_{o1n}-id_{n}N^{(1)}_{e1n})
\end{equation}
\begin{equation}
H_{1}=\frac{-k_{1}}{\omega \mu_{1}}\sum_{n=1}^{\infty} E_{n}(d_{n}M^{(1)}_{e1n}+ic_{n}N^{(1)}_{o1n})
\end{equation}
\begin{equation}
E_{2}=\sum_{n=1}^{\infty} E_{n}[f_{n}M^{(1)}_{o1n}-ig_{n}N^{(1)}_{e1n}+v_{n}M^{(2)}_{o1n}-iw_{n}N^{(2)}_{e1n}]
\end{equation}
\begin{equation}
H_{2}=\frac{-k}{w\mu_{2}}\sum_{n=1}^{\infty} E_{n}[g_{n}M^{(1)}_{e1n}+if_{n}N^{(1)}_{o1n}+w_{n}M^{(2)}_{e1n}+iv_{n}N^{(2)}_{o1n}]
\end{equation}
\begin{equation}
E_{3}=\sum_{n=1}^{\infty}E_{n}(ia_{n}N^{(3)}_{e1n}-b_{n}M^{(3)}_{o1n})
\end{equation}
\begin{equation}
H_{3}=\frac{k}{\omega \mu}\sum_{n=1}^{\infty} E_{n}(ib_{n}N^{(3)}_{o1n}+a_{n}M^{(3)}_{e1n})
\end{equation}

  \begin{equation}
  M_{o1n}=\cos\phi \pi_{n}(cos\theta)z_{n}(\rho)\hat{e_{\theta}}-sin\phi \tau_{n} (cos\theta)z_{n}(\rho)\hat{e_{\phi}}
 \end{equation}

   \begin{equation}
  M_{e1n}=-\sin\phi \pi_{n}(cos\theta)z_{n}(\rho)\hat{e_{\theta}}-cos\phi \tau_{n} (cos\theta)z_{n}(\rho)\hat{e_{\phi}}
 \end{equation}

 \begin{equation}
    N_{o1n}=\sin\phi \thinspace n(n+1) \thinspace  sin\theta  \thinspace \pi_{n}(cos\theta)\frac{z_{n}(\rho)}{\rho}\hat{e_{r}} +sin\phi \thinspace \tau_{n} (cos\theta)\frac{[\rho z_{n}(\rho)]'}{\rho}\hat{e_{\theta}}+cos\phi  \thinspace \pi_{n}(cos\theta)\frac{[\rho z_{n}(\rho)]'}{\rho}\hat{e_{\phi}}
 \end{equation}

 \begin{equation}
  N_{e1n}=\cos\phi \thinspace n(n+1) \thinspace sin\theta  \thinspace \pi_{n}(cos\theta)\frac{z_{n}(\rho)}{\rho}\hat{e_{r}} +cos\phi  \thinspace \tau_{n} (cos\theta)\frac{[\rho z_{n}(\rho)]'}{\rho}\hat{e_{\theta}}-sin\phi  \thinspace \pi_{n}(cos\theta)\frac{[\rho z_{n}(\rho)]'}{\rho}\hat{e_{\phi}}
 \end{equation}\\

 Where $i= \sqrt{-1}$; $E_{n}=i^nE_{0}\frac{2n+1}{n(n+1)}$; $\omega$ is the angular frequency; $k$ and $k_{1}$ are wave number in the sphere given by
$\frac{2\pi m}{\lambda_{0}}$  $\frac{2\pi m_{1}}{\lambda_{0}}$ respectively; $\lambda_{0}$  is the wavelength in vacuum, $m_{1}$ and $m_{2}$ are the refractive indices
of the core and coating relative to the surrounding medium; $\mu$, $\mu_{1}$ and $\mu_{2}$ are the permeabilities of surrounding medium, core, coating respectively;
$\pi_{n}$ and $\tau_{n}$ are angle-dependent functions given by $\frac{P^{(1)}_{n}}{sin\theta}$ and $\frac{dP^{(1)}_{n}}{d\theta}$ respectively, where $P^{(1)}_{n}$ is the
associated Legendre functions of the first kind of the degree n and order 1; $\hat{e_{r}}$, $\hat{e_{\theta}}$, $\hat{e_{\phi}}$ are unit vectors in spherical coordinates; $\theta$
is the scattering angle; `r' is the distance from the center of the coated sphere and $a_{n}$, $b_{n}$, $c_{n}$, $d_{n}$, $f_{n}$, $g_{n}$, $v_{n}$, $w_{n}$ are the spherical mode coefficients.

Superscripts appended to M and N in Eqs. (5-12) denotes the kind of spherical Bessel function ${z_{n}(\rho)}$ is:
\begin{enumerate}

\item (1) denotes $j_{n}(\rho)$,  which is defined as $\sqrt{\frac{\pi}{2\rho}}j_{n+\frac{1}{2}}$, where $j_{n+\frac{1}{2}}$ is the Bessel function of first kind.
\item (2) denotes $y_{n}(\rho)$,which is defined as $\sqrt{\frac{\pi}{2\rho}}y_{n+\frac{1}{2}}$, where $y_{n+\frac{1}{2}}$ is the Bessel function of second kind.
\item (3) denotes $h^{(1)}_{n}(\rho)$, known as spherical Hankel function, which is defined as $j_{n}(\rho) + iy_{n}(\rho)$.
\end{enumerate}

%  After applying the boundary conditions for a sphere with inner radius `a' and outer radius `b'.
%  \begin{align}
% \ (E_{2}-E_{1})\times\hat{e_{r}}= 0  &&  (H_{2}-H_{1})\times\hat{e_{r}}= 0                 &&  where,  r=a
% \end{align}
%  \vspace{-1.5em}
% \begin{align}
% \ (E_{3}+E_{i}-E_{2})\times\hat{e_{r}}= 0   &&   \ (H_{3}+H_{i}-H_{2})\times\hat{e_{r}}= 0   &&  where,  r=b
% \end{align}
% 
%  we get following equation in coefficients $a_{n}$, $b_{n}$, $c_{n}$, $d_{n}$, $f_{n}$, $g_{n}$, $v_{n}$, $w_{n}$.
%   \begin{equation}
%  \ f_{n}m_{1}\psi_{n}(m_{2}x)-v_{n}m_{1}\chi_{n}(m_{2}x)-c_{n}m_{2}\psi_{n}(m_{1}x)=0
%  \end{equation}
%  \begin{equation}
%  \ w_{n}m_{1}\chi'_{n}(m_{2}x)-g_{n}m_{1}\psi'_{n}(m_{2}x)+d_{n}m_{2}\psi_{n}(m_{1}x)=0
%  \end{equation}
%  \begin{equation}
%  \ v_{n}\mu_{1}\chi'_{n}(m_{2}x)-f_{n}\mu_{1}\psi'_{n}(m_{2}x)+c_{n}\mu_{2}\psi'_{n}(m_{1}x)=0
%  \end{equation}
% \begin{equation}
% g_{n}\mu_{1}\psi_{n}(m_{2}x)-w_{n}\mu_{1}\chi_{n}(m_{2}x)-d_{n}\mu_{2}\psi_{n}(m_{1}x)=0
% \end{equation}
% \begin{equation}
% \ m_{2}\psi'_{n}(y)-a_{n}m_{2}\xi'_{n}(y)-g_{n}\psi'_{n}(m_{2}y)+w_{n}\chi'_{n}(m_{2}y)=0
% \end{equation}
% \begin{equation}
% \ m_{2}b_{n}\xi_{n}(y)-m_{2}\psi_{n}(y)+f_{n}\psi_{n}(m_{2}y)-v_{n}\chi_{n}(m_{2}y)=0
% \end{equation}
% \begin{equation}
% \ \mu_{2}\psi_{n}(y)-a_{n}\mu_{2}\xi_{n}(y)-g_{n}\mu\psi_{n}(m_{2}y)+w_{n}\mu\chi_{n}(m_{2}y)=0
% \end{equation}
% \begin{equation}
%  \ b_{n}\mu_{2}\xi'_{n}(y)-\mu_{2}\psi'_{n}(y)+f_{n}\mu\psi'_{n}(m_{2}y)-v_{n}\mu\chi'_{n}(m_{2}y)=0
% \end{equation}
% % \vspace{0.5cm}
 As given by Eq (7), we studied nanoshells illuminated by plane electromagnetic waves. The final solutions for mode coefficients $a_{n}$, $b_{n}$, $c_{n}$, $d_{n}$, $f_{n}$, $g_{n}$,
 $v_{n}$, $w_{n}$ have been derived using boundary conditions given in \cite{22}

 \begin{equation}
 A_{n}=\frac{m_{2}\psi_{n}(m_{2}x)\psi'_{n}(m_{1}x)-m_{1}\psi'_{n}(m_{2}x)\psi_{n}(m_{1}x)}
 {m_{2}\chi_{n}(m_{2}x)\psi'_{n}(m_{1}x)-m_{1}\chi'_{n}(m_{2}x)\psi_{n}(m_{1}x)}
 \end{equation}
\begin{equation}
 B_{n}=\frac{m_{2}\psi_{n}(m_{1}x)\psi'_{n}(m_{2}x)-m_{1}\psi_{n}(m_{2}x)\psi_{n}(m_{1}x)}
 {m_{2}\chi'_{n}(m_{2}x)\psi_{n}(m_{1}x)-m_{1}\psi'_{n}(m_{1}x)\chi_{n}(m_{2}x)}
\end{equation}
\begin{equation}
a_{n}=\frac{\psi_{n}[\psi'_{n}(m_{2}y)-A_{n}\chi'_{n}(m_{2}y)]-m_{2}\psi'_{n}(y)[\psi_{n}(m_{2}y)-A_{n}\chi_{n}(m_{2}y)]}
 {\xi_{n}(y)[\psi'_{n}(m_{2}(y)-A_{n}\chi'_{n}(m_{2}y)]-m_{2}\xi'_{n}(y)[\psi_{n}(m_{2}y)-A_{n}\chi_{n}(m_{2}y)]}
 \end{equation}
 \begin{equation}
 b_{n}=\frac{m_{2}\psi_(y)[\psi'_{n}(m_{2}y)-B_{n}\chi'_{n}(m_{2}y)]-\psi'_{n}(y)[\psi_{n}(m_{2}y)-B_{n}\chi_{n}(m_{2}y)]}
 {m_{2}\xi_{n}(y)[\psi'_{n}(m_{2}(y)-B_{n}\chi'_{n}(m_{2}y)]-\xi'_{n}(y)[\psi_{n}(m_{2}y)-A_{n}\chi_{n}(m_{2}y)]}
 \end{equation}
 
 Where, $x=ka$ , $y=kb$, Ricacati-Bessel functions are given as $\chi_{n}(\rho)=-\rho y_{n}(\rho)$,
$\psi_{n}(\rho)=\rho j_{n}(\rho)$ and $\xi_{n}(\rho)=\rho h^{(1)}_{n}(\rho)$ and $\chi'_{n}(\rho)$, $\psi'_{n}(\rho)$, $\xi'_{n}(\rho)$ represents differentiation with respect to the
argument in the parenthesis. The numerical evaluations of cross-sections using the above Lorenz-Mie formalism \cite{22, 36, 37} were also verified and complemented
by numerical volume integral methods such as discrete dipole approximation. \cite{38} All the simulations were made with water, as an ambient medium of permittivity 1.77. These results of 
internal polarization are also used in elucidating the physics in the previous section.

 let $D_{x}=\frac{\psi'_{n}(m_{2}x)}{\psi_{n}(m_{2}x)}$
 $D_{y}=\frac{\psi'_{n}(m_{2}y)}{\psi_{n}(m_{2}y)}$

 \begin{equation}
  c_{n}=\frac{D_{x}v_{n}\chi_{n}(m_{2}x)-v_{n}\chi'_{n}(m_{2}x)}
              {\psi'_{n}(m_{1}x)-D_{x}\left(\frac{m_{2}}{m_{1}}\right)\psi_{n}(m_{1}x)}
 \end{equation}
 \begin{equation}
  d_{n}=\frac{w_{n}m_{1}\chi'_{n}(m_{2}x)-m_{1}D_{x}w_{n}\chi_{n}(m_{2}x)}
             {m_{1}D_{x}\psi_{n}(m_{1}x)-m_{2}\psi'_{n}(m_{1}x)}
 \end{equation}
 \begin{equation}
  f_{n}=\frac{v_{n}\chi_{n}(y)(m_{2}y)+m_{2}\psi_{n}(y)-m_{2}b_{n}\xi_{n}(y)} {\psi_{n}(m_{2}y)}
 \end{equation}
 \begin{equation}
  g_{n}=\frac{w_{n}\chi_{n}(m_{2}x)+d_{n}\psi_{n}(m_{1}x)}{\psi_{n}(m_{2}x)}
 \end{equation}
 \begin{equation}
 v_{n}=\frac{\psi'_{n}(y)-b_{n}\xi'_{n}+D_{y}m_{2}b_{n}\xi_{n}(y)-D_{y}m_{2}\psi_{n}(y)}
             {D_{y}\chi_{n}(m_{2}y)-\chi'_{n}(m_{2}y)}
 \end{equation}
 \begin{equation}
  w_{n}=\frac{a_{n}m_{2}\xi'_{n}(y)+D_{y}\psi_{n}(y)-\frac{\psi'_{n}(m_{2}y)}{\psi_{n}(m_{2}y)}a_{n}\xi_{n}(y)-m_{2}\psi'_{n}(y)}
              {\chi'_{n}(m_{2}y)-D_{y} \chi_{n}(m_{2}y)}
 \end{equation}

\begin{acknowledgments}
Srishti Garg thanks Arun. I for his support in graphics.
\end{acknowledgments}

\end{document}